\documentclass[11pt,a4paper]{article}
\usepackage{amsmath}
\usepackage{amssymb}
\usepackage{amsfonts}
\usepackage{amsthm} 
\usepackage{url}
\usepackage{graphicx}
\usepackage[utf8]{inputenc}
\usepackage{times} 

\newtheorem{Thm}{Theorem}

\newtheorem{Prob}[Thm]{Problem}

\newcommand{\readset}{{\mathcal R}}

\begin{document}

\title{Sensitive Long-Indel-Aware Alignment of Sequencing Reads}
\author{Tobias Marschall and Alexander Schönhuth\\[2ex]
Centrum Wiskunde \& Informatica, Amsterdam, Netherlands\\
{\tt \{tm|as\}@cwi.nl}}

\maketitle
\thispagestyle{empty}


\addtocounter{page}{-1}

\section{Introduction}
\label{sec.intro}
The tremdendous  advances in high-throughput sequencing technologies have made population-scale sequencing as performed in the 1000~Genomes project~\cite{1000genomes2010} and the Genome of the Netherlands project\footnote{\url{http://www.nlgenome.nl/}} possible.
In order to keep up with ever-increasing sequencing speeds, research on fast read mapping algorithms was indicated and led to substantial advances in the last decade.
Refer to \cite{Li2010c} for a review.
Today, most widely-used read aligners such as BWA \cite{Li2009a} and bowtie \cite{Langmead2009,Langmead2012} use a Burrows-Wheeler transform based index, allowing for fast and space-effient search for alignments seeds, which are then extended to full alignments of the given reads.

At the same time, next-generation sequencing has allowed genom-wide discovery of variations beyond single-nucleotide polymorphisms (SNPs), in particular of structural variations (SVs) like deletions, insertions, duplications, translocations, inversions, and even more complex rearrangements.
Algorithms for the discovery of SVs have been reviewed in \cite{Medvedev2009,Alkan2011}.
There are four main paradigms for SV discovery:
(1)~using the local coverage to detect deletions and estimate copy numbers, e.g.~\cite{Campbell2008,Chiang2009,Alkan2009,Sudmant2010,Yoon2009,Abyzov2011};
(2)~employing insert size statistics to call insertions and deletions, e.g.~\cite{Korbel2009,Hormozdiari2009,Chen2009,Lee2009,Sindi2009,Quinlan2010,Marschall2012a};
(3)~aligning reads containing an SV breakpoint, that is, allowing larger indels in alignments, e.g.~\cite{Mills2006,Ye2009,Emde2012};
(4)~(local) assembly in breakpoint regions. 
Recently, methods combining two of the above ideas have been developed \cite{Rausch2012,Zhang2012}. 

Furthermore, standard read aligners such as BWA~\cite{Li2009a}, bowtie2~\cite{Langmead2012}, and Stampy~\cite{Lunter2011} are able to map reads containing indels of limited length, usually up to a length of 50 base pairs.
Correctly placing indels is difficult, especially when only processing one read at a time.
Therefore, re-alignment procedures that simultaneously consider all reads mapping to a locus have been proposed.
The most widely used such method is part of the Genome Analysis Toolkit (GATK)~\cite{McKenna2010,DePristo2011}.
Although this strategy improves alignments of reads containing indels, it cannot make up for the lacking ability of current read mappers to detect long indels ($>$50bp).

In order to estimate the probability that an alignment is wrong, read mappers consider the set of alternative alignments.
When plausible alternative alignments are found, the probability of the primary aligment being correct is decreased.
To robustly estimate this probability, often referred to as mapping quality, it is thus crucial not to miss any relevant alternative alignment.
Furthermore, a statistically sound \emph{error model} is necessary to assess the set of alignments for a given read.
Most current read mappers use affine gap costs, which is a rather rough approximation of the true conditions in the human genome.

Based on the above observations, we design a read aligner with special emphasis on the following properties:
\begin{enumerate}
 \item \label{itm:sensitivity} high sensitivity, i.e.\ find all (reasonable) alignments,
 \item \label{itm:indels} ability to find (long) indels,
 \item \label{itm:scores} statistically sound alignment scores,
 \item \label{itm:runtime} runtime fast enough to be applied to whole genome data.
\end{enumerate}

\section{Approach}
Let $\readset$ be a set of paired-end reads where each read has sequenced ends of length $\ell$.
Formally, $\readset\subset \Sigma^\ell\times\Sigma^\ell$.
For current Illumina devices, $\ell$ is between 100 and 250\,bp.
In the following, we describe a 4-step procedure to map one single paired-end read $(S_1,S_2)\in\readset$ to a reference genome, as illustrated by Figure~\ref{fig:split-map-overview}.

\subsubsection*{Step 1: Global Anchor Search}
\begin{figure}[t]
\begin{center}
\includegraphics[width=\textwidth]{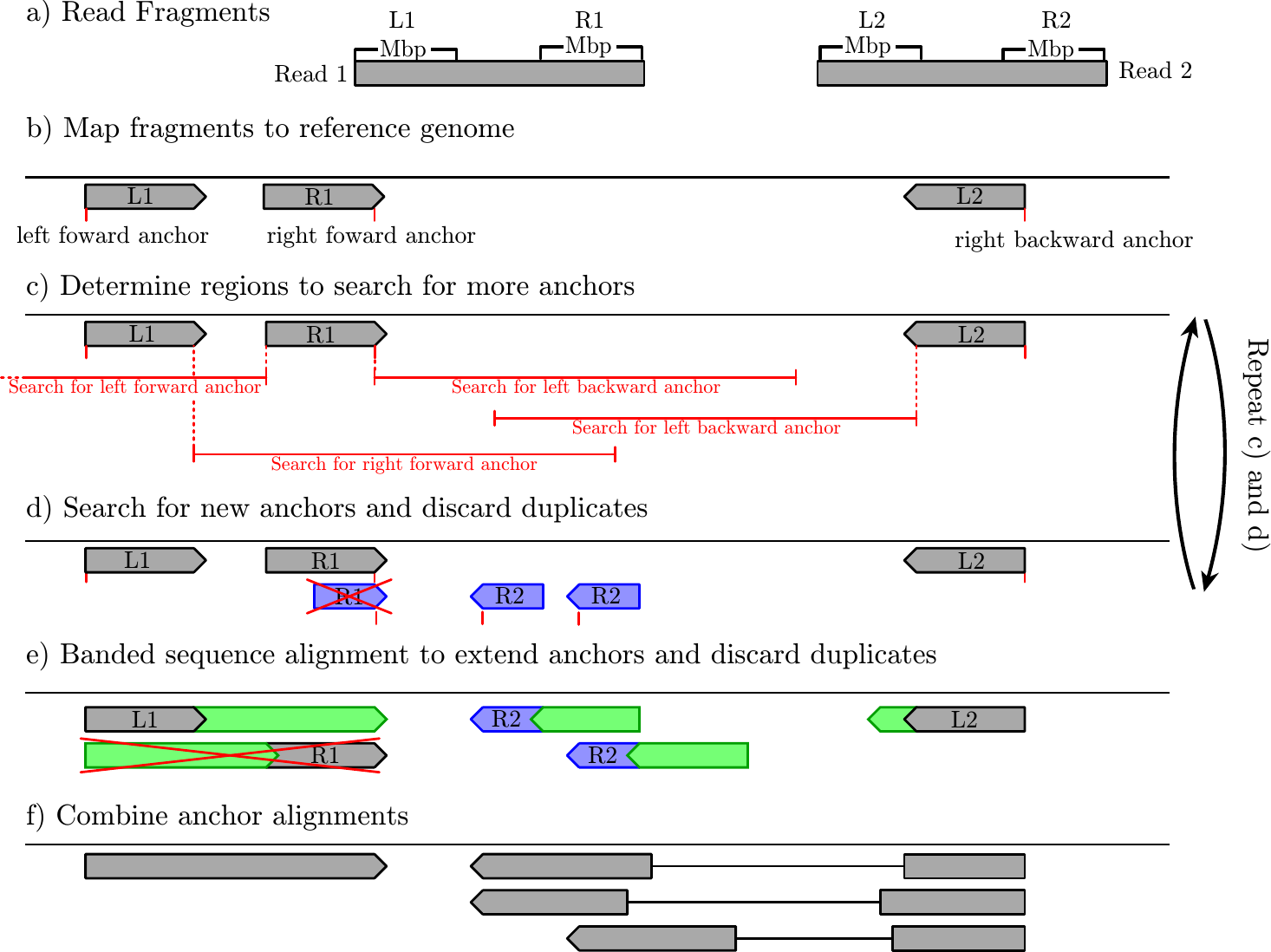}
\end{center}
\caption{Overview of the read mapping procedure.}\label{fig:split-map-overview}
\end{figure}

At first, we determine all genomic loci the read pair might map to.
These loci will later be subject to a more detailed analysis.
For this step, we make use of a standard Burrows-Wheeler-transform based read aligner as follows.
From each read of the pair $(S_1,S_2)$, we extract a length-$M$ prefix and suffix: we set $L1=S_1[1,M]$, $R1=S_1[\ell-M,\ell]$, $L2=S_2[1,M]$, and $R2=S_2[\ell-M,\ell]$, see Fig.~\ref{fig:split-map-overview}a.
These four fragments are then mapped by BWA~\cite{Li2009a} in single-end mode, allowing up to 25 alignments per fragment, see Fig.~\ref{fig:split-map-overview}b.
Any other read mapper can be used in this step, but we opt for BWA as it is mature and fast.
If BWA indicates (in the X0 and X1 tags) that number of alignments for a fragment exceeds 25, we treat that fragment as if it was unmapped.
This avoid any biases, as only the alignments of fragments for which BWA reported all possible alignments are used as anchor points for aligning the rest of the read pair.
The fragment length $M$ should thus be chosen such that most fragments do no produce more than 25 alignments.
In practice, setting $M=50$ works well.
The probability that all four fragments generated from a pair of reads cannot be mapped is small.
Or, in other words, it is unlikely that all four fragments are very ambiguous or contain indels BWA cannot handle.
Therefore, we will assume that the true locus of at least on of the four fragments is among the alignments returned by BWA.

Assuming the fragments $L1$ and $R1$ aligns to the forward strand and fragments $L2$ and $R2$ to the backward strand, we call the leftmost and rightmost alignment positions of fragment $L1$ and $R1$ the \emph{left forward anchor} and \emph{right forward anchor} (positions marked in red Fig.~\ref{fig:split-map-overview}b).
Analogously, we call the rightmost and leftmost alignment position of fragments $L2$ and $R2$ the \emph{right backward anchor} and \emph{left forward anchor}, as shown in Fig.~\ref{fig:split-map-overview}b.
Therefore, the anchors define the candidates for start and end positions of the alignments of a complete read.

\subsubsection*{Step 2: Local Anchor Search}
Our next goal is to determine additional anchor points, placing emphasis on sensitivity.
We assume that for any given fragment the correct alignment might not be among the set of alignments returned by the standard read mapper due to heavily repetitive areas, sequencing errors, excessive SNPs, structural variations, or a combination thereof.
As mentioned above, we only assume that there is at least one correct alignment among the alignments of the four fragments, implying that it is sufficient to restrict the search to the regions near fragment alignments in order to find the correct alignment of the whole read pair.
When searching genome wide, we had to choose $M$ large enough to avoid an excessive amount of false positives.
Now, we can restrict the search for additional anchors to regions neighboring anchors alignments produced by BWA,
Therefore, the search space is significantly smaller and we can search for shorter anchors without producing many false positives.

In detail, the local anchor search works as follows.
For each type of anchor we define regions to search for additional anchors.
For a left forward (backward) anchor, we search for right forward (backward) anchors to its right and for a right forward (backward) anchor, we search for left forward (backward) anchors to its left.
That ensures that we can find additional anchors belonging to the same read end.
To also find additional anchors for the respective mate read, we search for left backward anchor in the region left of a right forward anchor and search for right forward anchors in the region left of left backward anchors.
Selection of regions for local anchor search is shown in Fig.~\ref{fig:split-map-overview}c.
The search is done by means of an extended shift-and algorithm that finds all occurrences of a pattern within a given edit distance~\cite{Wu1992}.
This algorithm is a bit-parallel and thus fast implementation of a non-deterministic finite automaton. 
In practice, locally searching all regions closer than 2000\,bp to anchor alignments for anchors of length 14 allowing an edit distance of up to two yields good results.
Search regions with overlap, for example the regions for backward anchor search in Fig.~\ref{fig:split-map-overview}c, are merged and searched only once.
The newly found anchors (show in blue in Fig.~\ref{fig:split-map-overview}d) give rise to new regions to be searched.
We repeat this local anchor search procedure three times; ignoring regions already searched.
In case already known anchor points are (re-)found, we keep the original anchor and discard the new one.
The rationale for doing three rounds of local anchor search is the following.
Consider the (extreme) scenario in which only one of the four fragments, say L1, could be mapped by BWA.
Then, anchors for R1, L2, and R2 can be found in anchor search rounds one, two, and three, respectively.

\subsubsection*{Step 3: Partial Alignments from Anchor Points}
Next, each anchor is extended by a \emph{banded sequence alignment} as far as possible without exceeding a cost threshold~\cite{Chao1992}.
We explain banded sequence alignments in the next paragraph where we also discuss the used alignment cost scheme.
In the meantime, Figure~\ref{fig:split-map-overview}e shows anchors (gray and blue) that have been extended (green part).
Some anchors can be extended to an alignment of the full read (anchor L1 on the left).
If that is not possible, the partial alignment table for the respective prefix of the read end is kept to be joined with another partial alignment in Step 4.
Note that two anchors can give rise to the same alignment, see Figure~\ref{fig:split-map-overview}e.
We therefore discard duplicates before proceeding further.

When imposing a cost threshold on a pairwise sequence alignment with linear gap costs, the cost threshold and the gap costs determine how far a valid alignment can deviate from the main diagonal of the dynamic programming table.
For a cost threshold $t$ and gap costs $g$, only the $\lfloor t/g\rfloor$ diagonals above and below the main diagonal can contain entries not exceeding~$t$.
Therefore, it is sufficient to compute this part---or band---of the alignment table.
All alignments are based on ``phred''-like costs.
That is, an integer cost of $c$ is equivalent to a probability of $10^{-c/10}$ that the alignment is correct.
We use a cost threshold of $t=\text{115}$.
The cost of a mismatch equals the base quality as reported by the sequencing machine.
For indel costs, we proceed in two phases.
Here, in Step 3, we assign a cost of 30 to each 1\,bp indel and use linear gap-costs.
Later, in Step 5, when empirical distributions of insertion and deletion sizes are available, we recalibrate alignment scores to reflect these.

\subsubsection*{Step 4: Joining Partial Alignments}
Finally, for each extended fragment, we iterate over all pairs of left and right anchor alignments that are not too far apart (default: at most 1000\,bp).
For each such pair, we check whether the two (partial) alignment tables can be combined into one split-read alignment by introducing one (possibly larger) insertion or deletion, jumping from one table into the other.
This method is inspired by \emph{chaining}, which is a well-studied technique for aligning sequences based on known common fragments~\cite{Chao1995,Morgenstern2002}.
Recently, a similar technique for combining anchors to obtain split alignments was used in~\cite{Emde2012}.
One such large indel incurs a fixed cost (default: 25). 
In Step~5, these cost will be recalibrated in the same way as all indels contained in the anchor alignments.
There might be different possible ways of combining the two partial alignments, differing in the location and/or length of the indel.
For equivalent indels, we only retain the leftmost one.
For non-equivalent indels we retain all corresponding combined alignments that do not exceed the cost of the best alignment by more then a constant (default: 29).

To sum up, the result after this step are sets of alignments for both read ends, where each alignment can be either regular (obtained in Step~3) or a split alignment generated by joining two partial alignments.
Moreover, each alignment comes with an associated cost that can be interpreted as the probability that this alignment is correct.

\subsubsection*{Step 5: Recalibrating Alignment Scores}
In the final step, alignment scores of individual alignments and of \emph{pairs of alignments} are recalibrated.
To this end, we use empirical distributions of insert sizes, insertion sizes, and deletion sizes.
These distributions are obtained by considering all successfully aligned read pairs for which all four fragements L1, R1, L2, and R2 had exactly one anchor point.
Note that this is an even stronger criterion than requiring alignments to be unique.

During recalibration, all indel costs are replaced by the costs derived from the respective emprical distribution.
For each read pair $(S_1,S_2)$, we now compute scores for all pairs of alignment for $S_1$ and $S_2$.
The probability for a pair of aligments is the product of the probabilities for the two alignments and the (empirical) probability of the resulting insert size.

\section{Results}
\label{sec.results}
\subsection{Benchmark dataset}
For benchmarking, we use Venter's genome as described previously~\cite{Marschall2012a}.
All SNPs and structural variations~\cite{Levy2007} were inserted back into the reference genome to generate two alleles of each chromosome.
Then, 2$\times$100\,bp reads were simulated to 30-fold coverage according to Illumina error profiles as provided with the read simulator SimSeq \cite{Earl2011}.
Fragment sizes (including read ends) are sampled from a normal distribution with mean 500  and standard deviation 15.
We emphasize that this dataset reflects a true, diploid human genome and therefore the spatial distribution of SNPs and structural variations is realistic.
In combination with the realistic Illumina error profiles, this ensures that the difficulty of this benchmark dataset matches a dataset generated by an actual sequencer.

\begin{table}
\caption{Overview of reads by type. 
We only consider chromosome 22 from which a total of 10\,354\,336 reads have been generated.
Reads in insertion/deletion categories are those reads containing \emph{exactly one indel} of the respective type.
Reads containing more indels or reads ending in an insertion are subsumed in category \emph{Other}.
}\label{tab:read_overview}
\begin{center}
\begin{tabular}{lrr}
\hline
Read type & Count & Percentage \\\hline
No indels &  10\,090\,856 & 97.46 \\
Short deletion (1--20\,bp) & 96\,294 & 0.93 \\
Short insertion (1--20\,bp) & 102\,349 & 0.99 \\
Midsize deletion (21--50\,bp) & 3\,428 & 0.03 \\
Midsize insertion (21--50\,bp) & 2\,003 & 0.02 \\
Long deletion ($>$50\,bp) & 2\,083 & 0.02 \\
Other & 57\,323 & 0.55 \\\hline
\end{tabular}
\end{center}
\end{table}

During read simulation, we record where each read originated from and store its ``true'' alignment to the reference genome.
We classify each read according to whether its true alignment contains indels as summarized in Table~\ref{tab:read_overview}.
In the following, we will focus on the reads containing indels as these are difficult to align; reads without any indel do not pose major problems in terms of finding their alignment(s).

\subsection{Comparison to Other Read Mappers}
We run BWA~\cite{Li2009a}, Bowtie2~\cite{Langmead2012}, Stampy~\cite{Lunter2011}, and our approach on the set of reads generated from chromosome 22.
Furthermore, we postprocess all resulting alignments using the GATK~\cite{McKenna2010,DePristo2011} and report results with and without postprocessing.
Here, postprocessing consists of marking duplicates and realigning reads at indel loci.
Results are presented separately for each of the read categories shown in Table~\ref{tab:read_overview}.
For each category, we classify alignments according to a ``correctness hierarchy'' as follows.
An alignment is called \emph{fully correct} when it exactly matches the true alignment.
If that is not the case but start and end position match, that is, the number of inserted/deleted characters was correctly detected but indels where misplaced, we report the alignment as having \emph{correct end points}.
When an aligner anchored the read correctly but (for instance) failed to recognize an indel, it is reported as \emph{one correct end point}.
If both end points are wrong but the start position is not further than 5000\,bp away from the true start position, we classify the alignment as \emph{wrong but near}.
Otherwise, it is classified as \emph{wrong}.
Tables \ref{tab:short_del}, \ref{tab:short_ins}, \ref{tab:mid_del}, \ref{tab:mid_ins}, and \ref{tab:long_del} show performance rates on reads containing exactly one short deletion, short insertion, midsize deletion, midsize insertion, and long deletion, respectively.
Each read mapper was allowed to report up to 25 alignments per read, one of which it marks as \emph{primary}.
All tables contain two different types of results.
First, only primary alignments are considered.
Second, out of all alternative alignment, the most favorable one is picked, i.e.\ the one that matches the true alignment best.
This second statistics allow to assess whether misalignments are due to lacking sensitivity or due to wrong ranking of alignments.
Reads missing to 100~percent were unmapped according to the aligner.

\section{Discussion}
Our approach performs similar to existing read mappers on short indels from 1 to 20\,bp.
For midsize indels, which most readmapper can process, it has clear advantages over all tested competitors.
Even for long deletions, the performance of the presented approach is good; for 63.2\,\% of all reads containing long deletions, the correct alignment was among the returned alignments.
No other read mapper makes a significant contribution (or is intended to) in this length range.

\begin{table}
\caption{Performance on reads containing a \emph{short deletion} (1-20\,bp) and no other indels.
For each table cell, two percentages are given: the first represents the fraction of reads falling into that category, the second gives the cumulative fraction of reads in this or a higher category.
Numbers in bold mark the best cumulative percentage in the respective row.
}\label{tab:short_del}
\begin{center}
\begin{tabular}{lr@{ / }rr@{ / }rr@{ / }rr@{ / }r}
\hline
 & \multicolumn{2}{c}{New} & \multicolumn{2}{c}{BWA} & \multicolumn{2}{c}{Bowtie2} & \multicolumn{2}{c}{Stampy} \\\hline
\multicolumn{9}{l}{\textbf{Primary alignment only}}\\
        Fully correct & 69.4  &           69.4 &   67.5 &           67.5 & 71.4 &  \textbf{71.4} & 52.8 &           52.8 \\
   Correct end points &  5.7  &  \textbf{75.1} &    2.4 &           69.9 &  1.7 &           73.1 & 20.7 &           73.5 \\
One correct end point & 22.0  &           97.1 &   28.5 &           98.4 & 25.8 &           98.9 & 25.6 &  \textbf{99.1} \\
       Wrong but near &  0.2  &           97.4 &    0.4 &           98.7 &  0.1 &           99.0 &  0.1 &  \textbf{99.2} \\
                Wrong &  0.5  &           97.9 &    0.8 &           99.5 &  0.8 &           99.7 &  0.8 & \textbf{100.0} \\[.5em]
\multicolumn{9}{l}{\textbf{Most favorable alignment}}\\
        Fully correct & 75.8  &  \textbf{75.8} &   68.4 &           68.4 & 72.0 &           72.0 & 52.8 &           52.8 \\
   Correct end points &  7.5  &  \textbf{83.3} &    2.5 &           70.9 &  1.7 &           73.7 & 20.7 &           73.5 \\
One correct end point & 14.4  &           97.7 &   27.6 &           98.5 & 25.9 &  \textbf{99.6} & 25.6 &           99.1 \\
       Wrong but near &  0.0  &           97.8 &    0.4 &           98.8 &  0.0 &  \textbf{99.6} &  0.2 &           99.2 \\
                Wrong &  0.1  &           97.9 &    0.7 &           99.5 &  0.1 &           99.7 &  0.8 & \textbf{100.0} \\[.5em]
\multicolumn{9}{l}{\textbf{Primary alignment only, after GATK processing}}\\
        Fully correct &  75.8 &          75.8 & 76.9 &          76.9 & 73.2 &          73.2 & 80.8 &  \textbf{80.8} \\
   Correct end points &   3.8 &          79.6 &  2.2 &          79.1 &  1.7 &          74.9 &  2.6 &  \textbf{83.5} \\
One correct end point &  17.5 &          97.1 & 19.2 &          98.4 & 23.9 &          98.9 & 15.6 &  \textbf{99.1} \\
       Wrong but near &   0.2 &          97.4 &  0.4 &          98.7 &  0.1 &          99.0 &  0.1 &  \textbf{99.2} \\
                Wrong &   0.5 &          97.9 &  0.8 &          99.5 &  0.8 &          99.7 &  0.8 &  \textbf{100.0}\\[.5em]
\multicolumn{9}{l}{\textbf{Most favorable alignment, after GATK processing}}\\
        Fully correct &  79.9 &          79.9 & 77.1 &          77.1 & 73.8 &          73.8 & 80.8 &  \textbf{80.8} \\
   Correct end points &   4.9 & \textbf{84.8} &  2.2 &          79.3 &  1.7 &          75.5 &  2.6 &           83.5 \\
One correct end point &  13.0 &          97.7 & 19.2 &          98.5 & 24.1 & \textbf{99.6} & 15.6 &           99.1 \\
       Wrong but near &   0.0 &          97.8 &  0.4 &          98.8 &  0.0 & \textbf{99.6} &  0.1 &           99.2 \\
                Wrong &   0.1 &          97.9 &  0.7 &          99.5 &  0.1 &          99.7 &  0.8 & \textbf{100.0} \\\hline
\end{tabular}
\end{center}
\end{table}

\begin{table}
\caption{Performance on reads containing a \emph{short insertion} (1-20\,bp) and no other indels.
For each table cell, two percentages are given: the first represents the fraction of reads falling into that category, the second gives the cumulative fraction of reads in this or a higher category.
Numbers in bold mark the best cumulative percentage in the respective row.
}\label{tab:short_ins}
\begin{center}
\begin{tabular}{lr@{ / }rr@{ / }rr@{ / }rr@{ / }r}
\hline
 & \multicolumn{2}{c}{New} & \multicolumn{2}{c}{BWA} & \multicolumn{2}{c}{Bowtie2} & \multicolumn{2}{c}{Stampy} \\\hline
\multicolumn{9}{l}{\textbf{Primary alignment only}}\\
        Fully correct & 68.2 &          68.2 & 64.1 &          64.1 & 71.0 & \textbf{71.0} & 53.0 &           53.0 \\
   Correct end points & 11.9 & \textbf{80.1} &  8.0 &          72.2 &  7.8 &          78.7 & 24.4 &           77.4 \\
One correct end point & 16.8 &          96.9 & 25.9 &          98.0 & 19.9 &          98.6 & 21.4 &  \textbf{98.7} \\
       Wrong but near &  0.2 &          97.1 &  0.5 &          98.6 &  0.1 &          98.7 &  0.1 &  \textbf{98.9} \\
                Wrong &  0.6 &          97.7 &  1.0 &          99.6 &  0.9 &          99.7 &  1.1 & \textbf{100.0} \\[.5em]
\multicolumn{9}{l}{\textbf{Most favorable alignment}}\\
        Fully correct & 72.2 & \textbf{72.2} & 65.0 &          65.0 & 71.7 &          71.7 & 53.0 &           53.0 \\
   Correct end points & 14.1 & \textbf{86.3} &  8.3 &          73.3 &  7.8 &          79.5 & 24.4 &           77.4 \\
One correct end point & 11.1 &          97.5 & 24.8 &          98.2 & 20.0 & \textbf{99.4} & 21.4 &           98.7 \\
       Wrong but near &  0.0 &          97.5 &  0.5 &          98.7 &  0.0 & \textbf{99.4} &  0.1 &           98.9 \\
                Wrong &  0.2 &          97.7 &  0.9 &          99.6 &  0.2 &          99.7 &  1.1 & \textbf{100.0} \\[.5em]
\multicolumn{9}{l}{\textbf{Primary alignment only, after GATK processing}}\\
        Fully correct & 72.6 &          72.6 & 72.4 &          72.4 & 72.6 &          72.6 & 76.9 &  \textbf{76.9} \\
   Correct end points & 10.1 &          82.7 &  7.6 &          80.0 &  7.8 &          80.4 &  9.4 &  \textbf{86.3} \\
One correct end point & 14.2 &          96.9 & 18.1 &          98.0 & 18.3 &          98.6 & 12.5 &  \textbf{98.8} \\
       Wrong but near &  0.2 &          97.1 &  0.5 &          98.6 &  0.1 &          98.7 &  0.1 &  \textbf{98.9} \\
                Wrong &  0.6 &          97.7 &  1.0 &          99.6 &  0.9 &          99.7 &  1.1 & \textbf{100.0} \\[.5em]
\multicolumn{9}{l}{\textbf{Most favorable alignment, after GATK processing}}\\
        Fully correct & 75.6 &          75.6 & 72.5 &          72.5 & 73.3 &          73.3 & 76.9 &  \textbf{76.9} \\
   Correct end points & 11.8 & \textbf{87.4} &  7.7 &          80.2 &  7.8 &          81.1 &  9.4 &           86.3 \\
One correct end point & 10.0 &          97.5 & 18.0 &          98.2 & 18.3 & \textbf{99.4} & 12.5 &           98.8 \\
       Wrong but near &  0.0 &          97.5 &  0.5 &          98.7 &  0.0 & \textbf{99.4} &  0.1 &           98.9 \\
                Wrong &  0.2 &          97.7 &  0.9 &          99.6 &  0.2 &          99.7 &  1.1 & \textbf{100.0} \\ \hline
\end{tabular}
\end{center}
\end{table}

\begin{table}
\caption{Performance on reads containing a \emph{midsize deletion} (21-50\,bp) and no other indels.
For each table cell, two percentages are given: the first represents the fraction of reads falling into that category, the second gives the cumulative fraction of reads in this or a higher category.
Numbers in bold mark the best cumulative percentage in the respective row.
}\label{tab:mid_del}
\begin{center}
\begin{tabular}{lr@{ / }rr@{ / }rr@{ / }rr@{ / }r}
\hline
 & \multicolumn{2}{c}{New} & \multicolumn{2}{c}{BWA} & \multicolumn{2}{c}{Bowtie2} & \multicolumn{2}{c}{Stampy} \\\hline
\multicolumn{9}{l}{\textbf{Primary alignment only}}\\
        Fully correct & 38.0 & \textbf{38.0} & 11.1 &          11.1 &  0.0 &           0.0 &  0.9 &           0.9\\
   Correct end points &  2.6 & \textbf{40.6} &  0.4 &          11.5 &  0.0 &           0.0 & 16.3 &          17.2\\
One correct end point & 47.3 &          87.9 & 78.1 &          89.6 & 80.3 &          80.3 & 77.0 & \textbf{94.2}\\
       Wrong but near &  0.9 &          88.8 &  4.6 &          94.2 &  2.0 &          82.3 &  2.6 & \textbf{96.8}\\
                Wrong &  1.0 &          89.8 &  1.8 &          96.0 &  2.5 &          84.8 &  2.5 & \textbf{99.3}\\[.5em]
\multicolumn{9}{l}{\textbf{Most favorable alignment}}\\
        Fully correct & 66.5 & \textbf{66.5} & 11.1 &          11.1 &  0.0 &           0.0 &  0.9 &           0.9\\
   Correct end points &  9.3 & \textbf{75.7} &  0.4 &          11.5 &  0.0 &           0.0 & 16.3 &          17.2\\
One correct end point & 13.0 &          88.7 & 79.5 &          91.0 & 82.5 &          82.5 & 77.0 & \textbf{94.2}\\
       Wrong but near &  0.1 &          88.8 &  3.2 &          94.2 &  0.3 &          82.8 &  2.8 & \textbf{97.0}\\
                Wrong &  1.0 &          89.8 &  1.8 &          96.0 &  2.0 &          84.8 &  2.3 & \textbf{99.3}\\[.5em]
\multicolumn{9}{l}{\textbf{Primary alignment only, after GATK processing}}\\
        Fully correct & 48.7 & \textbf{48.7} & 17.9 &          17.9 &  0.0 &           0.0 & 44.9 &          44.9\\
   Correct end points &  1.6 & \textbf{50.3} &  0.7 &          18.6 &  0.0 &           0.0 &  2.0 &          46.9\\
One correct end point & 37.6 &          87.9 & 71.2 &          89.8 & 80.3 &          80.3 & 47.5 & \textbf{94.5}\\
       Wrong but near &  0.9 &          88.8 &  4.4 &          94.2 &  2.0 &          82.3 &  2.4 & \textbf{96.8}\\
                Wrong &  1.0 &          89.8 &  1.8 &          96.0 &  2.5 &          84.8 &  2.5 & \textbf{99.3}\\[.5em]
\multicolumn{9}{l}{\textbf{Most favorable alignment, after GATK processing}}\\
        Fully correct & 70.9 & \textbf{70.9} & 17.9 &          17.9 &  0.0 &           0.0 & 44.9 &          44.9\\
   Correct end points &  6.4 & \textbf{77.3} &  0.7 &          18.6 &  0.0 &           0.0 &  2.0 &          46.9\\
One correct end point & 11.4 &          88.7 & 72.6 &          91.2 & 82.5 &          82.5 & 47.5 & \textbf{94.5}\\
       Wrong but near &  0.1 &          88.8 &  3.0 &          94.2 &  0.3 &          82.8 &  2.6 & \textbf{97.0}\\
                Wrong &  1.0 &          89.8 &  1.8 &          96.0 &  2.0 &          84.8 &  2.3 & \textbf{99.3}\\\hline
\end{tabular}
\end{center}
\end{table}

\begin{table}
\caption{Performance on reads containing a \emph{midsize insertion} (21-50\,bp) and no other indels.
For each table cell, two percentages are given: the first represents the fraction of reads falling into that category, the second gives the cumulative fraction of reads in this or a higher category.
Numbers in bold mark the best cumulative percentage in the respective row.
}\label{tab:mid_ins}
\begin{center}
\begin{tabular}{lr@{ / }rr@{ / }rr@{ / }rr@{ / }r}
\hline
 & \multicolumn{2}{c}{New} & \multicolumn{2}{c}{BWA} & \multicolumn{2}{c}{Bowtie2} & \multicolumn{2}{c}{Stampy} \\\hline
\multicolumn{9}{l}{\textbf{Primary alignment only}}\\
        Fully correct & 36.5 & \textbf{36.5} &  2.0 &           2.0 &  0.0 &           0.0 &  2.1 &            2.1\\
   Correct end points & 12.4 & \textbf{48.9} &  1.1 &           3.1 &  0.0 &           0.0 & 22.5 &           24.6\\
One correct end point & 30.7 &          79.6 & 73.9 &          77.1 & 50.9 &          50.9 & 56.3 &  \textbf{80.8}\\
       Wrong but near &  4.3 &          84.0 & 13.2 &          90.3 & 10.8 &          61.7 & 11.2 &  \textbf{92.1}\\
                Wrong &  2.7 &          86.7 &  5.5 &          95.8 &  9.7 &          71.4 &  7.9 & \textbf{100.0}\\[.5em]
\multicolumn{9}{l}{\textbf{Most favorable alignment}}\\
        Fully correct & 51.5 & \textbf{51.5} &  2.0 &           2.0 &  0.0 &           0.0 &  2.1 &            2.1\\
   Correct end points & 19.7 & \textbf{71.2} &  1.1 &           3.1 &  0.0 &           0.0 & 22.5 &           24.6\\
One correct end point & 13.0 & \textbf{84.2} & 74.0 &          77.1 & 60.4 &          60.4 & 56.3 &           80.8\\
       Wrong but near &  1.0 &          85.2 & 13.2 &          90.3 &  2.8 &          63.2 & 11.7 &  \textbf{92.5}\\
                Wrong &  1.5 &          86.7 &  5.5 &          95.8 &  8.2 &          71.4 &  7.4 & \textbf{100.0}\\[.5em]
\multicolumn{9}{l}{\textbf{Primary alignment only, after GATK processing}}\\
        Fully correct & 48.2 & \textbf{48.2} &  2.4 &           2.4 &  0.0 &           0.0 & 37.8 &           37.8\\
   Correct end points &  9.4 & \textbf{57.6} &  1.1 &           3.5 &  0.0 &           0.0 &  6.7 &           44.5\\
One correct end point & 22.3 &          79.9 & 73.5 &          77.1 & 51.1 &          51.1 & 37.1 &  \textbf{81.6}\\
       Wrong but near &  4.0 &          84.0 & 13.2 &          90.3 & 10.6 &          61.7 & 10.4 &  \textbf{92.1}\\
                Wrong &  2.7 &          86.7 &  5.5 &          95.8 &  9.7 &          71.4 &  7.9 & \textbf{100.0}\\[.5em]
\multicolumn{9}{l}{\textbf{Most favorable alignment, after GATK processing}}\\
        Fully correct & 60.4 & \textbf{60.4} &  2.4 &           2.4 &  0.0 &           0.0 & 37.8 &           37.8\\
   Correct end points & 14.9 & \textbf{75.2} &  1.1 &           3.5 &  0.0 &           0.0 &  6.7 &           44.5\\
One correct end point &  8.9 & \textbf{84.2} & 73.6 &          77.1 & 60.4 &          60.4 & 37.1 &           81.6\\
       Wrong but near &  1.0 &          85.2 & 13.2 &          90.3 &  2.8 &          63.2 & 10.9 &  \textbf{92.5}\\
                Wrong &  1.5 &          86.7 &  5.5 &          95.8 &  8.2 &          71.4 &  7.4 & \textbf{100.0}\\\hline
\end{tabular}
\end{center}
\end{table}

\begin{table}
\caption{Performance on reads containing a \emph{long deletion} ($>$50\,bp) and no other indels.
For each table cell, two percentages are given: the first represents the fraction of reads falling into that category, the second gives the cumulative fraction of reads in this or a higher category.
Numbers in bold mark the best cumulative percentage in the respective row.
}\label{tab:long_del}
\begin{center}
\begin{tabular}{lr@{ / }rr@{ / }rr@{ / }rr@{ / }r}
\hline
 & \multicolumn{2}{c}{New} & \multicolumn{2}{c}{BWA} & \multicolumn{2}{c}{Bowtie2} & \multicolumn{2}{c}{Stampy} \\\hline
\multicolumn{9}{l}{\textbf{Primary alignment only}}\\
        Fully correct &  19.3 & \textbf{19.3} &  0.0 &           0.0 &  0.0 &           0.0 &  0.0 &            0.0\\
   Correct end points &   1.7 & \textbf{21.1} &  0.0 &           0.0 &  0.0 &           0.0 &  0.6 &            0.6\\
One correct end point &  50.8 &          71.9 & 77.9 &          77.9 & 72.3 &          72.3 & 80.9 &  \textbf{81.6}\\
       Wrong but near &   6.6 &          78.5 & 18.3 &          96.3 & 16.2 &          88.6 & 16.1 &  \textbf{97.6}\\
                Wrong &   1.0 &          79.5 &  1.7 &          97.9 &  1.5 &          90.1 &  2.4 & \textbf{100.0}\\[.5em]
\multicolumn{9}{l}{\textbf{Most favorable alignment}}\\
        Fully correct &  51.6 & \textbf{51.6} &  0.0 &           0.0 &  0.0 &           0.0 &  0.0 &            0.0\\
   Correct end points &  11.6 & \textbf{63.2} &  0.0 &           0.0 &  0.0 &           0.0 &  0.6 &            0.6\\
One correct end point &  14.5 &          77.6 & 85.6 & \textbf{85.6} & 84.6 &          84.6 & 80.9 &           81.6\\
       Wrong but near &   0.9 &          78.5 & 11.3 &          96.9 &  4.7 &          89.3 & 16.1 &  \textbf{97.6}\\
                Wrong &   1.0 &          79.5 &  1.1 &          97.9 &  0.8 &          90.1 &  2.4 & \textbf{100.0}\\[.5em]
\multicolumn{9}{l}{\textbf{Primary alignment only, after GATK processing}}\\
        Fully correct &  26.6 & \textbf{26.6} &  0.0 &           0.0 &  0.0 &           0.0 &  4.6 &            4.6\\
   Correct end points &   1.4 & \textbf{28.0} &  0.0 &           0.0 &  0.0 &           0.0 &  0.0 &            4.6\\
One correct end point &  43.9 &          71.9 & 78.0 &          78.0 & 72.7 &          72.7 & 77.2 &  \textbf{81.8}\\
       Wrong but near &   6.6 &          78.5 & 18.3 &          96.3 & 15.8 &          88.6 & 15.8 &  \textbf{97.6}\\
                Wrong &   1.0 &          79.5 &  1.7 &          97.9 &  1.5 &          90.1 &  2.4 & \textbf{100.0}\\[.5em]
\multicolumn{9}{l}{\textbf{Most favorable alignment, after GATK processing}}\\
        Fully correct &  53.1 & \textbf{53.1} &  0.0 &           0.0 &  0.0 &           0.0 &  4.6 &            4.6\\
   Correct end points &  10.4 & \textbf{63.6} &  0.0 &           0.0 &  0.0 &           0.0 &  0.0 &            4.6\\
One correct end point &  14.0 &          77.5 & 85.6 & \textbf{85.6} & 84.8 &          84.8 & 77.2 &           81.8\\
       Wrong but near &   1.0 &          78.5 & 11.2 &          96.9 &  4.5 &          89.3 & 15.8 &  \textbf{97.6}\\
                Wrong &   1.0 &          79.5 &  1.1 &          97.9 &  0.8 &          90.1 &  2.4 & \textbf{100.0}\\\hline
\end{tabular}
\end{center}
\end{table}

\newpage
\bibliographystyle{plain}
\bibliography{misterl}

\end{document}